\documentclass[conference]{IEEEtran}
\usepackage[T1]{fontenc}
\usepackage[utf8]{inputenc}
\usepackage[USenglish]{babel}
\usepackage{hyperref}

\ifCLASSINFOpdf
  \usepackage[pdftex]{graphicx}
  \graphicspath{{img/}}
\else
\fi

\usepackage{stfloats}
\begin{document}
%
\title{Real-Time Redundancy for the 1.3 GHz\\ Master Oscillator of the European-XFEL}

\author{\IEEEauthorblockN{Bartosz Gąsowski, Tomasz Owczarek \\Krzysztof Czuba and Łukasz Zembala}
\IEEEauthorblockA{Institute of Electronic Systems\\
Warsaw University of Technology\\
Warsaw, Poland\\
Email: b.gasowski@elka.pw.edu.pl}
\and
\IEEEauthorblockN{Holger Schlarb}
\IEEEauthorblockA{Deutsches Elektronen-Synchrotron\\
Hamburg, Germany}}


%


\maketitle

\begin{abstract}
Many modern large-scale facilities, like European X-ray Free Electron Laser (E-XFEL), require precise synchronisation, often down to femtosecond level. Even a very short interruption or an excessive glitch in the reference signal might break the precise time relations between subsystems. In such event, a time-consuming resynchronization process is required that renders the facility not available for the users until it is completed. Therefore, such events are highly undesirable.

In this paper, we present an autonomous redundancy solution for the European-XFEL's master oscillator that will guarantee a continuous delivery of the high-quality reference signal even in case of most of the potential failures. The concept and implementation are presented, as well as results from testing in the laboratory environment.
\end{abstract}


%
\IEEEpeerreviewmaketitle

\section{Introduction}
Many modern large-scale facilities, like for example X-ray Free Electron Lasers, require synchronisation with minute accuracy often reaching down to femtosecond level. An example of such facility is European X-ray Free Electron Laser (European-XFEL)~\cite{xfel}, an over 3.4~km long complex which is located at Deutsches Elektronen-Synchrotron site (Hamburg, Germany).
At such scale of complexity achieving required accuracy results in time-consuming system setup and hence requirement for continuous operation.

\subsection{Synchronisation overview}
Synchronisation at such levels of accuracy is usually realised by means of delivery and use of a reference signal whose phase is well-defined and stable.
For the purpose of a simplified analysis, synchronisation issues can be split into short- and long-term stability of reference signal phase.

Short-term stability is considered in the terms of phase noise, often also expressed as integrated jitter. 
Most of the challenges in the construction of the master oscillator system come from the phase noise requirements.
Long-term phase stability is mainly a result of drifts of the electrical and/or optical lengths in the reference's distribution network (including cables as well as components).
Synchronisation system's main task is compensation of those drifts in order to maintain phase relations between synchronized nodes within required accuracy.

Long-term frequency stability is another matter as it mostly concerns the primary frequency reference inside the master oscillator.
With proper operation of the synchronisation system, stability of this reference dominates the overall frequency stability.

Reference signal is very often used for synthesis of local derived signals in LLRF and other systems.
Example of such signals are local oscillator (LO) and ADC clock signals, whose frequency and phase relations to the reference are well defined.
Synthesis of these signals usually employs circuits like phase-lock loops and frequency dividers, which require continuous reference signal for proper operation.

\subsection{Motivation}

Stringent performance requirements imposed on the synchronisation system, including the master oscillator, in turn impose stringent requirements on the components of these systems.
This very often reduces the available spectrum of solutions and parts to a narrow range.
Consequently, performance as an requirement that is clearly expressed in quantitative terms and has to be met, is prioritised over reliability.

In fact, some failures were already observed during development and test runs of European-XFEL's master oscillator, as well as in similar systems.
Failures can be permament (complete failure of some component) or intermittent (e.g. overheating caused by cooling system failure, cold joints, or EMI issues).
Therefore, they can result in a variety of reference signal's disturbances, ranging from complete loss of signal, to temporary power dips, to phase jumps.

When such disturbances occur in the input signal of a phase-lock loop or a frequency divider, events like cycle-slips can occur.
Subsequently, well-defined time relation between the input and output signals is no longer guaranteed.
Usually, frequency dividers that are used in such cases have a reset input which allow to synchronise them again.
However, in such a large and complex systems a race condition will probably occur between the reset and reference signals, and thus this solution will not guarantee deterministic behaviour.

What is more, Europen-XFEL employs complex mixed RF and optical system for distribution of the reference signal \cite{sync_ipac13} \cite{mlo_ipac17}. Effects of reference signal's disturbances on these systems were not yet analysed in detail. However, it cannot be excluded that after a disturbance the distribution system might require a warm-up-like period in order to return to desired long-term stability, introducing addidtional delays.

Thus, any interruptions or significant disturbances in the reference signal might break well-defined time relations between subsystems and result in a time-consuming process of resynchronisation and set-up.
At such scale of complexity recovery to femtosecond stability and set-up of the LLRF system  can easily take hours, if not days.
During this time facility is unavailable for the users which is highly undesirable.
In this paper we present an autonomous redundancy solution which aims to mitigate reliability issues in the source of the reference signal, the master oscillator.

\section{Concept}

\subsection{Overview}
Prior solutions for minimizing downtime of the master oscillator are limited to maintaining of a hot spare. It means that there is a continuously running spare copy of the system. In case a failure occurs in the main system, the facility is reconnected to the spare one. However, as the failure already occured and the reference signal has been disturbed or lost, issues outlined in the previous section still hold true. Our solution for European-XFEL aims to mitigate these issues.

\begin{figure*}[t]
\centering
\includegraphics*[width=0.8\textwidth,trim=0 0 0 0]{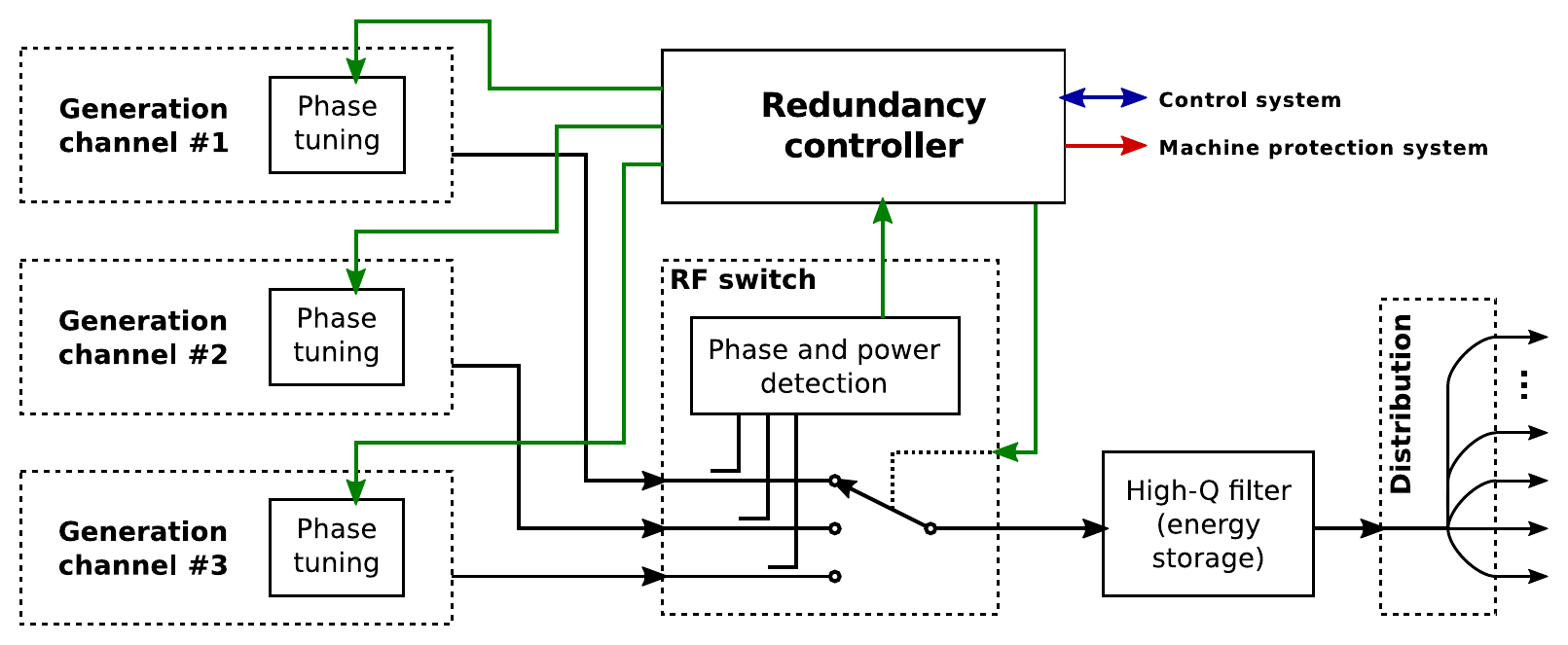}
\caption{Simplified block diagram of the European-XFEL's master oscillator system.}
\label{fig_block_diag}
\end{figure*}

An overview of the European-XFEL's master oscillator system is presented in fig.~\ref{fig_block_diag}, with emphasis on redundancy. 
The system consists of three identical and independent reference signal sources (generation channels, GC) and a redundancy subsystem. Each GC produces a very high quality 1.3 GHz reference signal (< 20~fs rms jitter, < $10^{-12}$ frequency stability, +41~dBm power) and is fully independent from the other two. More detailed information on the generation channels can be found in~\cite{mo_ipac14}.

The main objective of the system is to maintain a continuous and proper reference signal even in case of failure in one of the GCs, and minimise or smooth-out any disturbances that might be caused by such events. The concept is based on energy storage in a high quality-factor dielectric resonator filter, low-latency detection of phase and/or amplitude disturbances, and fast switching between generation channels. The filter delays failure propagation to the output, allowing for a brief period of time for reaction. 

Nominally, all three channels are equal and fully operational at all times and provide their output signals to the redundancy subsystem. However, only signal from one of the channels is passed further at a time—this channel is regarded as a current master, and the other two are treated as spares. If a potentially critical failure is detected in the current master channel, a switchover occurs and one of the former spares becomes a new master. Critical failure is understood as a failure that potentially renders output signal unusable and could result in losing synchronism in the European-XFEL, as explained in the previous section. Due to the properties of the filter, the total available time budget for failure detection and reaction is estimated to be approximately 200--300 nanoseconds.

\subsection{Failure Detection}
The system aims to maintain reference signal continuity even in case of disturbances caused by failures. All disturbances will manifest themselves in irregularities of the phase or the amplitude, or both of them. In other words, an actual issue are the discontinuities of the phase and amplitude.

Thus, the concept of failure detection is built around monitoring of the phase and amplitude of the reference signals. All three signals (from all generation channels) are continously monitored for disturbances, and based on that their validity is assessed. With this approach it is possible to reliably detect all important failures, where by important failure we understand one that causes unacceptable disturbances in the reference signal.
Example of a most likely "unimportant" failure is a degradation of the phase noise performance. Of course, an excessive phase noise is undesirable, but neither does it cause cycle-slips\footnote{Of course, if the phase noise increases enough to cause cycle-slips, then it is an unacceptable phase disturbance and should be considered as such.}, nor can it be measured reliably in a short time, especially at femtosecond levels. It such case the switchover can be done manually by the operator at a later time.

While monitoring of the amplitude is trivial, monitoring of the phase poses additional challenge. The amplitude can be easily measured as an absolute value. The phase, however, can only be measured in a relative manner.
Therefore, in our system the phase is measured between the generation channels, in pairs. If we consider only a single pair of sources, then incorrect value of measured phase would only indicate that failure occured, but it would not give information on which channel did actually fail.
In case of three sources, three phase measurements are available. And while the values of phase related to the failing source will change, the third value---measured between properly operating sources---will not.
Therefore, it is possible to use an approach similar to voting employed in digital circuits and infer which generation channel did fail.
Of course, it is assumed that it is very unlikely for two channels to fail at the same moment.

\begin{figure}[t]
\centering
\includegraphics*[width=\columnwidth,trim=0 0 0 0]{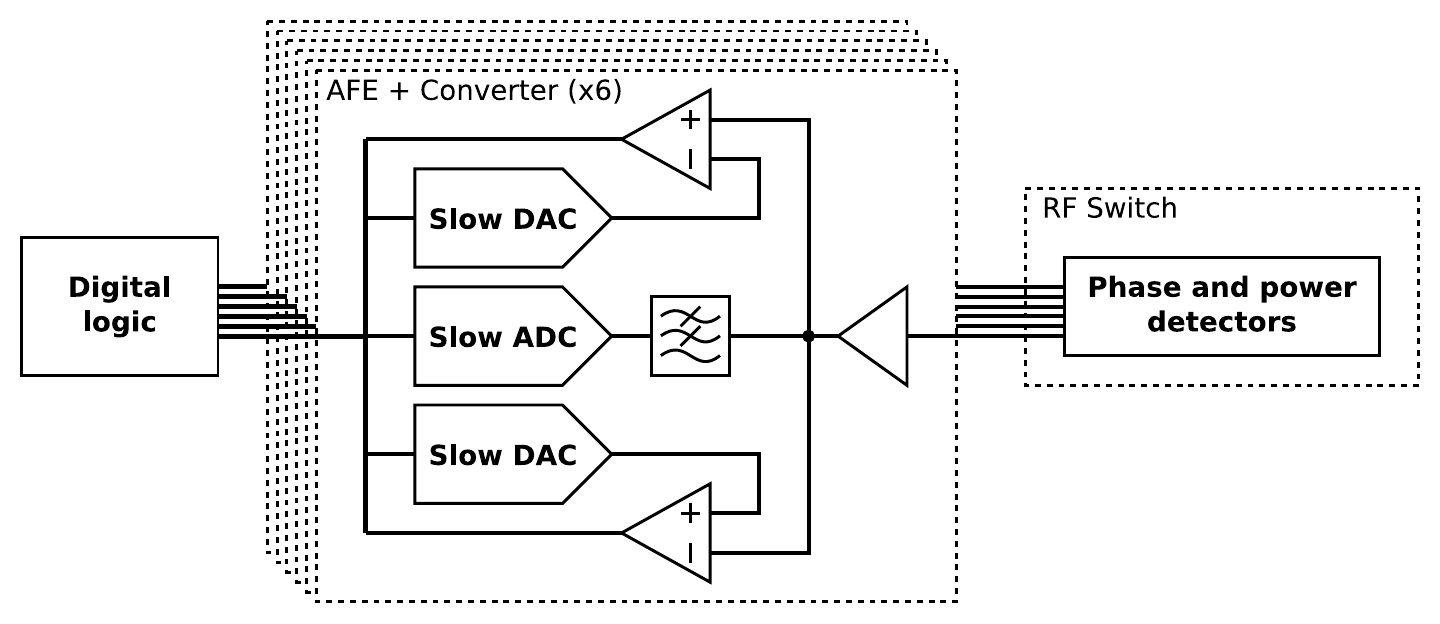}
\caption{Concept of low-latency monitoring with programmable window comparators.}
\label{fig_afe}
\end{figure}

For lowest latency, part of the system that is responsible for fast detection of failures processes signals from the phase and amplitude detectors primarily in the analogue domain.
Use of a fast ADC and digital domain processing was found to offer higher latency and increased design complexity.
The solution employs programmable window comparators in order to continously check if the values are withing expected range, as shown in Fig.~\ref{fig_afe}.
Out-of-range signals are then passed to the digital circuits for further processing, according to the description in the previous paragraph.
Slow DACs are used to control the thresholds of the comparators as desired. Slow ADC is also included in order to enable closed-loop control of the thresholds and analyze long-term changes, among others.

\subsection{Maintaining Reference Continuity}
When a failure is detected in the currently selected generation channel (a master channel), a quick switchover to another one is required in order to maintain continuity of the output signal. A three-way RF switch module is used to select one of the channels and to switch between them within 100 ns of a trigger signal. This module is required to handle reference signals with power levels reaching +41 dBm while offering reasonable insertion loss and high isolation. The concept of this module was already developed and presented earlier in~\cite{rfs_icalepcs2017}.

At the very output of the system a high quality-factor filter is inserted as a temporary RF energy storage. It sustains output power during intermittent signal interruptions, caused either by a failure or the switching itself. It also helps to smooth out any disturbances of the phase and the amplitude.

As switching between the channels should not induce excessive phase changes on its own, synchronisation (phase alignment) between the generation channels is required. Therefore, each GC has a phase tuning capability which is used to align their phases. The spare channels follow current master's phase, while the phase shifter of the master channel is always frozen as not to cause any excess jitter in the output signal. In the role of the phase shifters vector modulators are proposed, because their phase shifting range is not constrained and so they allow for continuous tracking of phase. The synchronisation loop shares the phase detectors and ADCs with detection of phase failures.

\section{Hardware Implementation}
The redundancy subsystem was implemented as a set of hadware modules.
An overwiev of each module is presented in the following subsections.

\subsection{RF Switch Module}
The RF switch module is a custom three-way RF switch circuit which has integrated phase and amplitude detectors. It is able to handle +43 dBm signals and switch between the channels within about 50 ns. Insertion loss has an acceptable value of about 2~dB, while isolation exceeds 80~dB.
The module is already developed and tested; the details were presented earlier in~\cite{rfs_icalepcs2017}.

\subsection{Filter}
The filter that was developed for this system is a dual dieletric resonator filter, characterized by a loaded Q-factor of about 4000. The Q-factor of the resonators themselves exceeds 10000. Insertion loss is below 1.5~dB and the filter is capable of handling signals with power exceeding 50~dBm. More detailed description of the filter's construction is available in~\cite{filterAA}.

Thanks to the relatively large Q-factor, this filter is able to sustain output power for few hundred nanoseconds. For example, in case the input power disappears, after 300~ns the output power will drop by about 3~dB. This gives rough estimation of time budget available for failure reaction path.

\subsection{Redundancy Controller}
\begin{figure}[t]
\centering
\includegraphics*[width=\columnwidth,trim=0 0 0 0]{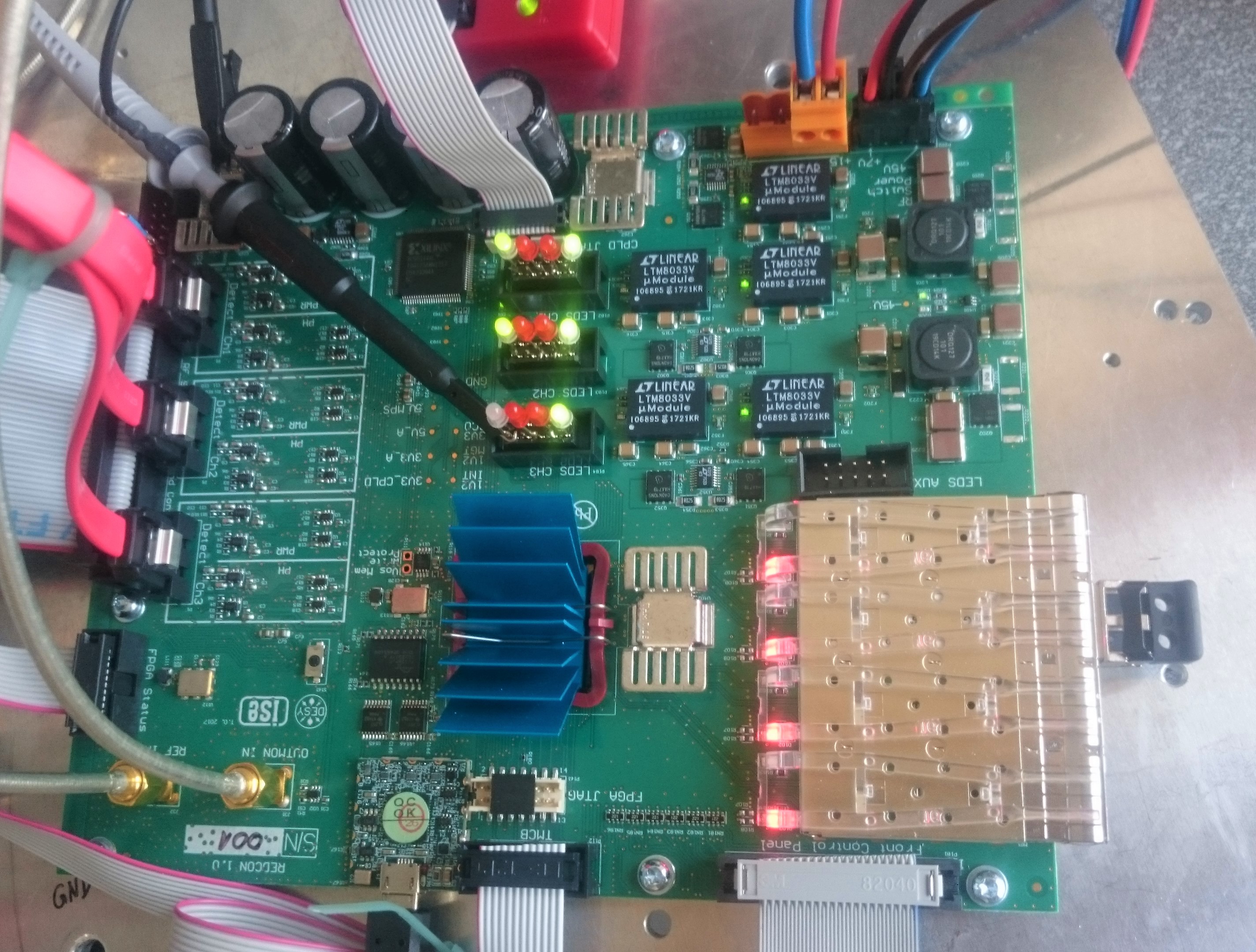}
\caption{Photograph of the redundancy controller.}
\label{fig_redcon}
\end{figure}

The redundancy controller is a central module of the system, controlling and supervising operation of the other modules. Manufactured device is shown in Fig.~\ref{fig_redcon}. Core components of this module are an analogue processing circuits, a CPLD, and an FPGA. The analogue processing circuits are implemented according to concept introduced in the previous section. Outputs of the comparators are then passed to the CPLD. The CPLD contains an asynchronous low-latency logic which is rensponsible for failure decoding, decision making and control of the RF switch module.
The logic is implemented with triple modular redundancy (TMR) for increased reliability.

The FPGA is responsible for general control, communication and supervision.
It handles channel synchronisation (phase alignment) through connected phase shifting modules, control of the thresholds for the window comparators, monitoring of system's state, and is an interface to the control system of the facility. The FPGA can be reprogrammed remotely without interruption in module's operation---separate CPLD circuit ensures that critical failure reaction paths are continously operational. Also, thanks to a built-in mechanism the FPGA will reprogram itself when it detects errors in its configuration memory, in particular those caused by a single event upset (SEU).

Additionally, the controller module contains curcuit for an auto-calibration of phase detector set-points. Because of large tolerance of the phase detector parameters (which themselves are within the RF switch module), set-point calibration is necessary for obtaining good accuracy of channel synchronisation.
Power supply sections and sources are redundant in order to ensure contonuous operation.

\subsection{Phase Shifter}

As a phase shifters DRTM-VM2LF vector modulator modules are used. These modules were developed primarily for use in European-XFEL's LLRF system based on mTCA.4 platform~\cite{vm_ipac13}. In case of our system these modules operate in a standalone mode with a custom firmware. Also, they are modified for better phase noise performance, i.e. analogue bandwidth of the baseband is reduced to a very narrow value of about 10~Hz.

The vector modulator modules are the only part of the redundancy system that is contained within the generation channels. In each generation channel the modulator module is placed just before the last synthesizer. It is required because of power levels: the last synthesizer incorporates high-power amplifier and the power level of its output signal reaches ca 41~dBm.
Additionally, it helps to significantly reduce influence of residual phase noise of the modulator.
The DRTM-VM2LF modules are connected with the redundancy controller via optical fibre links.

\section{Tests}

\subsection{Test Setup and Method}

\begin{figure}[t]
\centering
\includegraphics*[width=\columnwidth,trim=0 0 0 0]{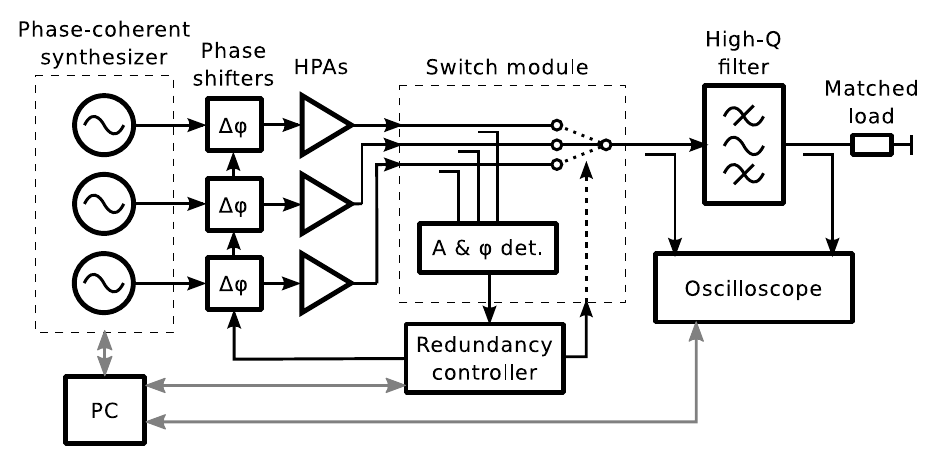}
\caption{Block diagram of the test setup.}
\label{fig_teststand}
\end{figure}

Block diagram of the test setup is presented in Fig.~\ref{fig_teststand}.
The generation channels are emulated by a multichannel phase-coherent synthesizer, a set of the vector modulator modules, and a set of microwave high power amplifiers (only for high power tests).
The amplifiers are currently not available and thus only tests with lower power levels were carried on with the full system. This is, however, enough to do a functional verification and prove correctnes of the concept. The modules that are in the path of the high power signal were tested earlier separately (the RF switch module up to 42~dBm, the filter up to 50~dBm). Therefore, no issues are expected during high power tests of the full system.

There are two proposed methods for an injection of artificial failures. One method is to inject them in the vector modulators. While this method offers a better control over the signal, it does also require a modified firmware. It might not be desirable, however, to use a different firmware during the tests than during the normal operation.
In the second method the failures are emulated within the multichannel synthesizer by changing its settings through a control interface. While this method offers more coarse control of the signals, it does not require the modified firmware and is simpler to implement.
Results presented in the next subsection were obtained with help of the second method.

Output signal (from before and after the filter) is recorded on a digital oscilloscope with a bandwidth of 4~GHz and sampling rate of 10 GHz, and then processed on a PC. Processing comprises an IQ demodulation with an averaging window of 100 samples (13 full periods of the 1.3~GHz signal) followed by a conversion of IQ values to the amplitude (magnitude) and phase (angle). 

\subsection{Test Results}

We present results that were obtained without a closed synchronisation loop (manual phase alignment) and with lower power signals (no high-power amplifiers).
These preliminary tests were meant to verify correctness of system's operation and to assess the latency of its reaction to the failures.
Two main types of failure were analysed: power loss and phase change.

\subsubsection{Power Loss}
\begin{figure}[t]
\centering
\includegraphics*[width=\columnwidth,trim=0 0 0 0]{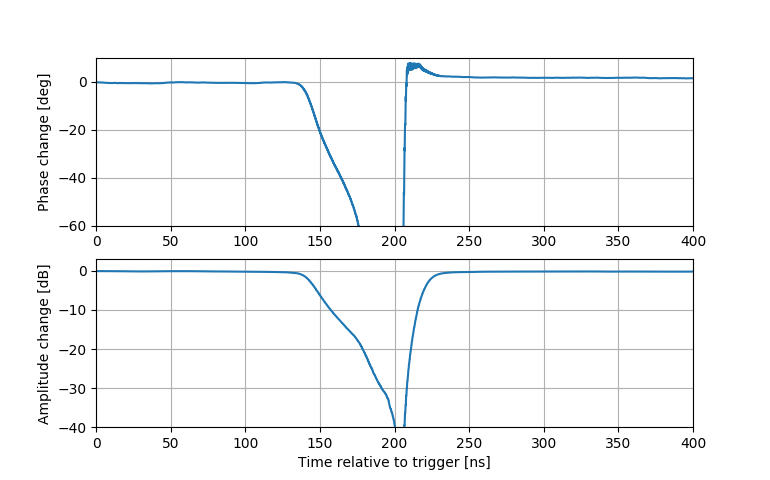}
\caption{Reaction to power loss---phase and amplitude of signal before the filter.}
\label{fig_powerdrop}
\end{figure}

Fig.~\ref{fig_powerdrop} shows phase and amplitude of the signal before filter in case of a power loss event.
It's clearly visible that at about 140~ns power started to decrease.
About 90~ns later a switchover to another channel is already completed.
Because of the synthesizer's behaviour the phase was also changing. However, in this case reaction was triggered first by the change of the amplitude.

\begin{figure}[t]
\centering
\includegraphics*[width=\columnwidth,trim=0 0 0 0]{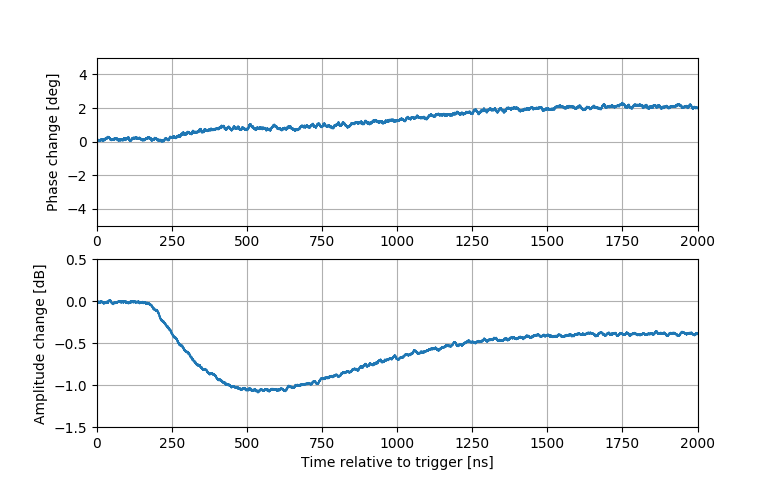}
\caption{Reaction to power loss---phase and amplitude of signal after the filter (system output).}
\label{fig_powerdrop_filter}
\end{figure}

Fig.~\ref{fig_powerdrop_filter} shows phase and anplitude in a similar event, but after the filter (at the system output).
Note the compressed time axis and stretched phase and amplitude axes. 
Both the amplitude and the phase change smoothly.
In the extremum the amplitude has decreased only by about 1~dB; however some amplitude unbalance between channels is also visible.
Phase change is mostly a result of channel alignment error (ca 2$^\circ$).

\subsubsection{Phase Change}
\begin{figure}[t]
\centering
\includegraphics*[width=\columnwidth,trim=0 0 0 0]{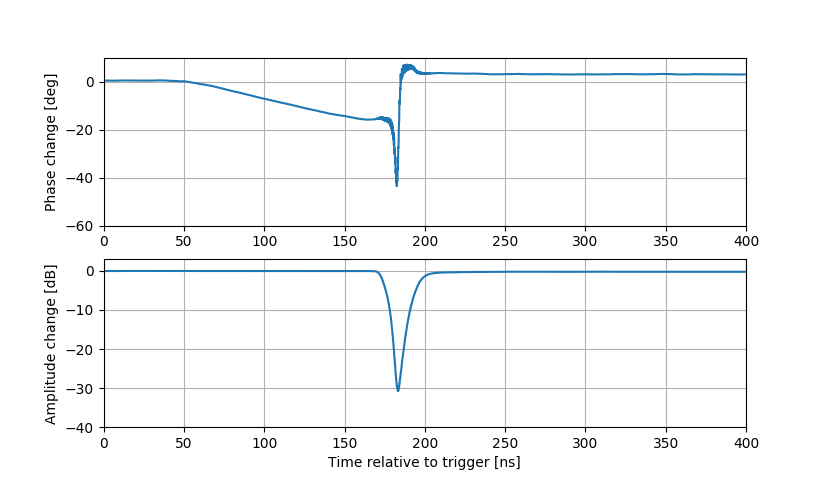}
\caption{Reaction to phase change---phase and amplitude of signal after the filter (system output).}
\label{fig_phasechange}
\end{figure}

Fig.~\ref{fig_phasechange} shows phase and amplitude of the signal before filter in case of a phase change.
The phase starts changing at about 50~ns. In this case failure detection latency is higher and the switchover to another channel is complete after about 150~ns. Increased latency is caused by two connected factors: slow phase change and a wide window in the window comparator.
The slower change occurs, the higher latency is acceptable though. On the other hand, the wide window is a result of increased noise of the phase detectors (due to lower power levels) and lack of the (calibrated) channel synchronisation.
In nominal conditions window shall be narrower and hence latency should decrease. The unbalance of the phase and the amplitude is similar as in the previous case.

\begin{figure}[t]
\centering
\includegraphics*[width=\columnwidth,trim=0 0 0 0]{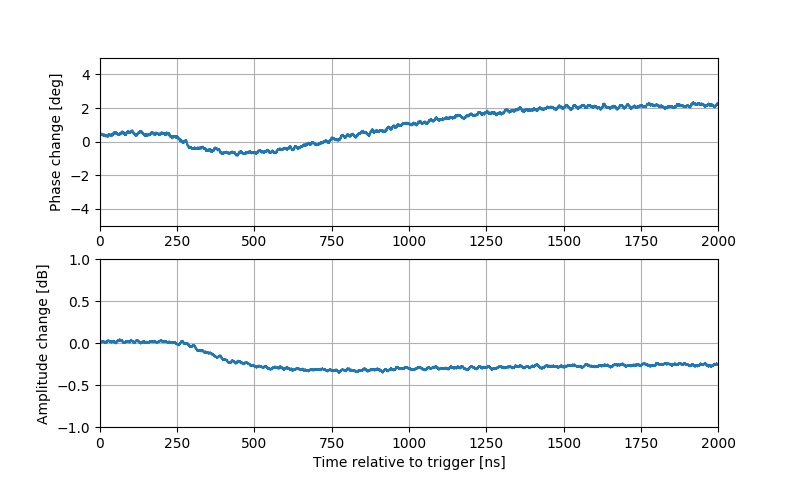}
\caption{Reaction to phase change---phase and amplitude of signal after the filter (system output).}
\label{fig_phasechange_filter}
\end{figure}

As previously, Fig.~\ref{fig_phasechange_filter} shows phase and amplitude after the filter (at the system output). In this case the phase is also changing smoothly.
The power almost do not change except as a result of the amplitude unbalance.

\section{Conclusions and further plans}
All hardware modules of the redundancy system for the European-XFEL's master oscillator are ready and tested. The test setup is partially completed and 
the tests of the whole system are currently in progress.
Preliminary tests have shown that concept of the redundancy is valid and it is possible to react in failure quickly enough---i.e. in less than 200~ns.
The high-Q filter properly smooths out distrubances ensuring continuity of the output signal.

Further firmware development is required to include missing functions, e.g the channel synchronisation and the auto-calibration.
After completion of both, the firmware and the teststand, the system will be tested thoroughly and then commisioned in the facility.


\section*{Acknowledgment}
Research supported by Polish Ministry of Science and Higher Education, founds for international co-financed projects for year 2017.



%

\end{document}